# Cosmology with Gamma-Ray Bursts Using k-correction

A. Kovács, Z. Bagoly, L. G. Balázs, I. Horváth, P. Veres

**Abstract**

In the case of Gamma-Ray Bursts with measured redshift, we can calculate the k-correction to get the fluence and energy that were actually produced in the comoving system of the GRB. To achieve this we have to use well-fitted parameters of a GRB spectrum, available in the GCN database. The output of the calculations is the comoving isotropic energy $E_{iso}$, but this is not the endpoint: this data can be useful for estimating the $\Omega_M$ parameter of the Universe and for making a GRB Hubble diagram using Amati's relation.

## 1 Introduction

Several papers present how to make k-corrections (e.g. [1]). We will only note here the principles and the most important considerations. A typical GRB spectrum has three main available parameters: peak energy ($E_{peak}$), low- and high energy spectral indices ($\alpha, \beta$). With the Band function [2] and the Cutoff power-law function (hereafter CPL) these are well-fitted. The motivation of k-correction is related to the fact that the energy distribution measured with gamma-ray detectors placed on satellites is not equal to the energy that the burst released in its comoving system. The goal is to correct the redshifted GRB spectrum. We will now show the main steps of this method and the meaning of k.

## 2 Theory of the k-correction

By definition, fluence is the integral of the Band/CPL function over the energy range in which the detector is responsive (see equation 1). If redshift is taken into account, the formula for bolometric energy has to be modified.

$$S_{[a,b]} = \int_a^b \Phi(E)\,\mathrm{d}E \qquad (1)$$

$$E_{[E_1,E_2]} = \frac{4\pi D_L^2}{(1+z)} S_{[E_{\min},E_{\max}]} \cdot \frac{S_{[E_1/(1+z),E_2/(1+z)]}}{S_{[E_{\min},E_{\max}]}} \qquad (2)$$

where $E_1 = 1$ keV and $E_2 = 10\,000$ keV, conventionally. The detectors do not measure the fluence between $[E_1/(1+z), E_2/(1+z)]$. All of them have an $[E_{\min}, E_{\max}]$ interval that they detect in, so the fluence should be corrected, as can be seen in equations (2) and (3). Hereafter $E_{[E_1,E_2]}$ is named $E_{iso}$.

$$E_{[E_1,E_2]} = \frac{4\pi D_L^2}{(1+z)} S_{[E_{\min},E_{\max}]} \cdot k \qquad (3)$$

Now we can see the meaning of k: it is a factor that multiplies the fluence, and therefore the energy. The most necessary parameters are also clear. The Gamma-Ray Burst Coordinate Network (GCN) provides a good database for the calculations. We only have to search for GRBs with measured redshift and decide whether the fit of the spectrum contains a peak energy parameter or not. This is important, because it is needed when working with Amati's relation [3]. We finally found 72 GRB samples, most of which are Konus-WIND data. However, some data is from other sources. Of course, the data of one GRB sample is not mixed. It is important to say some words about the error calculations. We used a somewhat unusual method: because each measured parameter in the GCN database has an error, we generated random Gaussian distributions where the mean value was the measured number, and the variance was the error. The k-corrections were made with these numbers from the tails of the curves, so we finally obtained a distribution for every GRB energy that has a mean value and a variance. We thus identified the error as the variance in each case. With this procedure the $\Delta E_{iso}$ errors are approximately one order less than the $E_{iso}$ energies.

## 3 Results of the k-corrections

We can clearly see from Figure 1 that the most frequented k value is approximately 1.2. Figure 2 shows the distribution of the isotropic energies. These energies are used when we want to test the Amati relation [4], which states that there is a correlation between $E_{peak}(1+z)$ ($E_{peak}$ in keV) and $\log_{10}(E_{iso})$ (see Figure 3). There are some outlier points. These are short GRBs (def.: $t_{90} < 2s$) which do not follow this relation. The stars refer to points using the CPL function, the squares are from the Band function. Our group wanted to test the trimodality of the GRB distribution. As mentioned above, short GRBs do not follow the Amati's relation, but the others can also be separated into groups in the Amati plane. Figure 4 shows these datapoints. There is an overlap between the intermediate ($2s < t_{90} < 10s$) and the long type groups (signed $*$), so the separation cannot be seen in our data. With more burst in the future it may be visible.

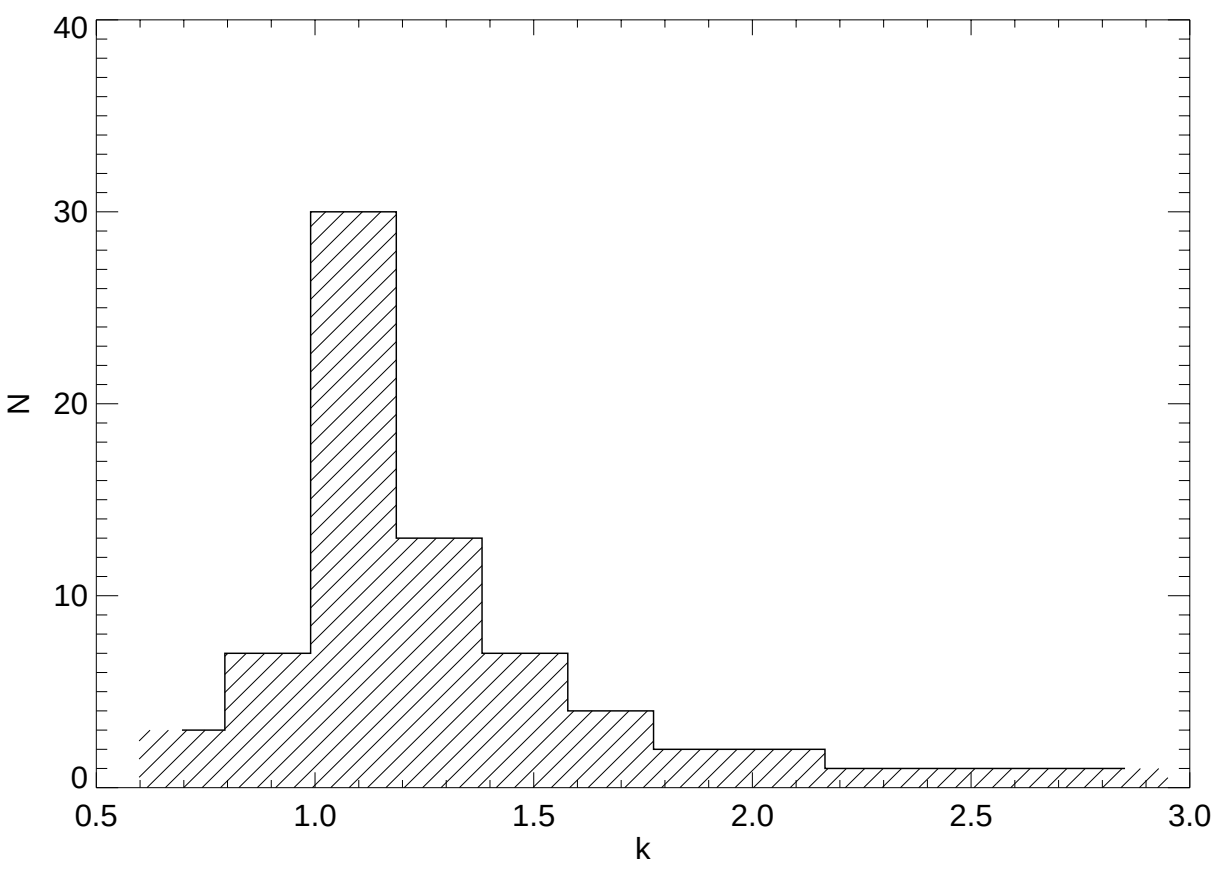

Fig. 1: Distribution of k

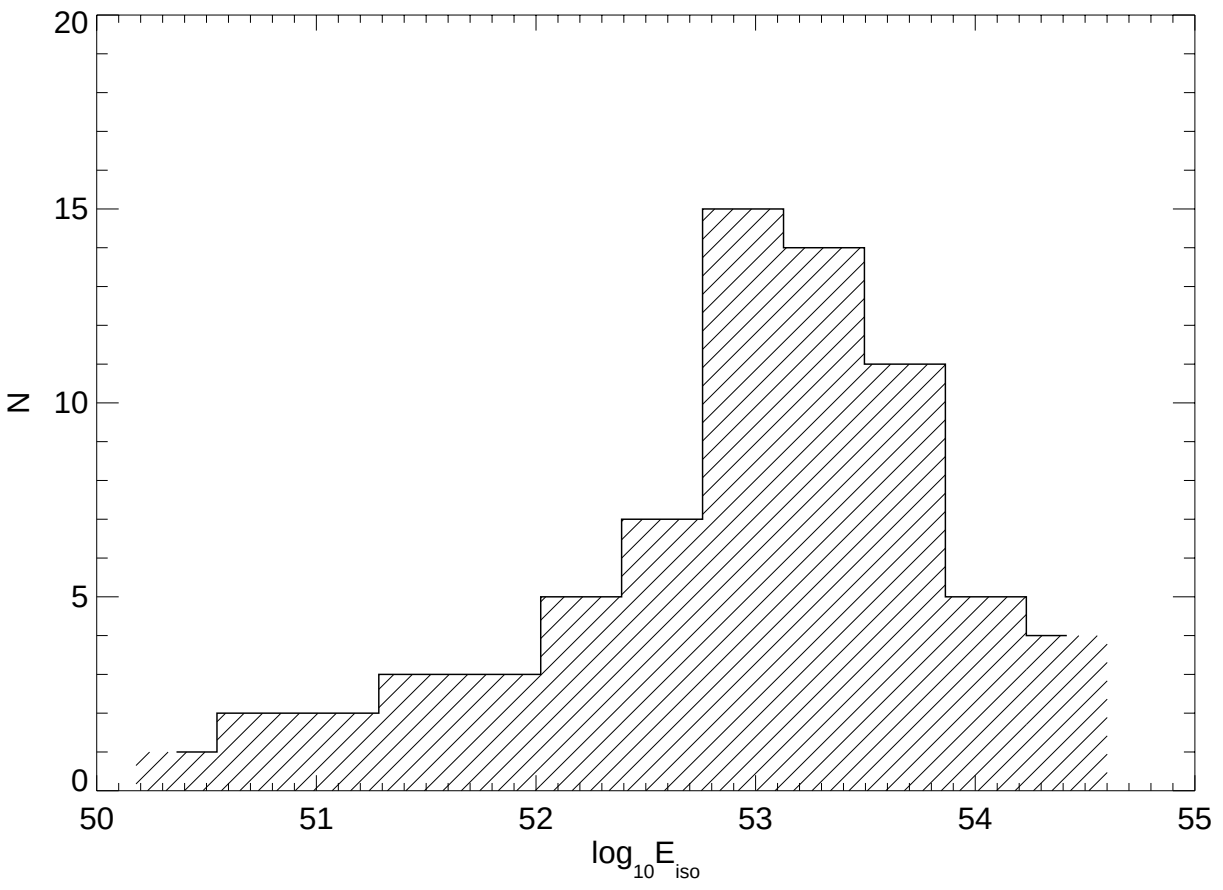

Fig. 2: Distribution of $\log_{10}(E_{iso})$, $E_{iso}$ is measured in ergs

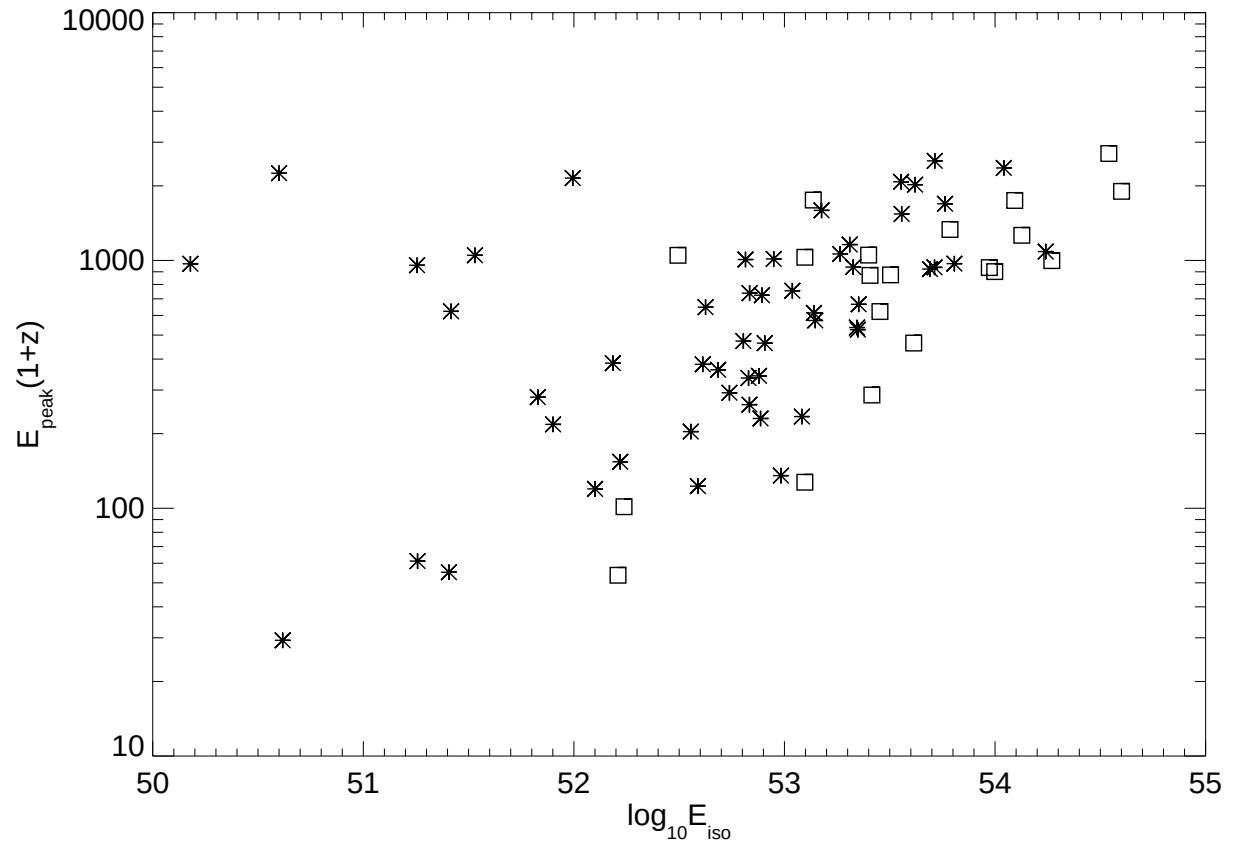

Fig. 3: Visualization of Amati's relation

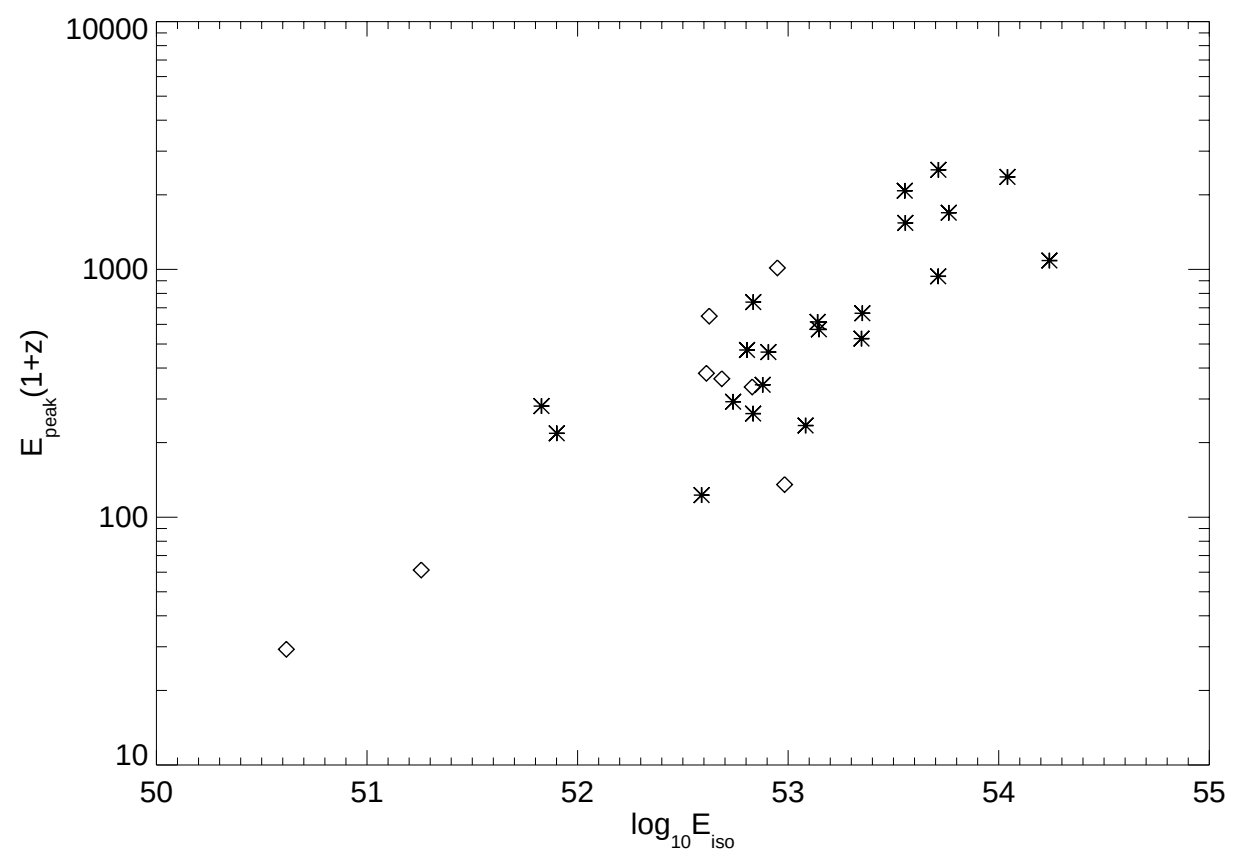

Fig. 4: The intermediate and the long GRB groups

# 4 Cosmological applications

We will now present the most interesting results that can be derived form the dataset, the Hubble diagram and an estimation of $\Omega_M$. The first interesting result is based on two important things: accelerating expansion of the Universe from Supernova Cosmology Project data [5] and earlier results using GRBs [6]. Our method was very simple, as we just wanted to make estimations. We were curious whether the time-dilatation effect is seen or not. Let us note the steps in this procedure: first of all, consider that the Amati relation is real and fit a straight line to the datapoints. After this, we can calculate $D_L$ using the $E_{iso}$ values related to the two parameters of the straight line. Finally, we ought to think that the points are exactly on the line. In this way we can get the luminosity distance in a different way, as written below in equation (4).

$$D_L = \left( \frac{E_{iso}(1+z)}{S_{[E_{min},E_{max}]} 4\pi k} \right)^{\frac{1}{2}} \tag{4}$$

The last step in this operation is to put the derived $D_L$ to the well-known distance modulus formula (see equation 5), and the final result is Figure 5.

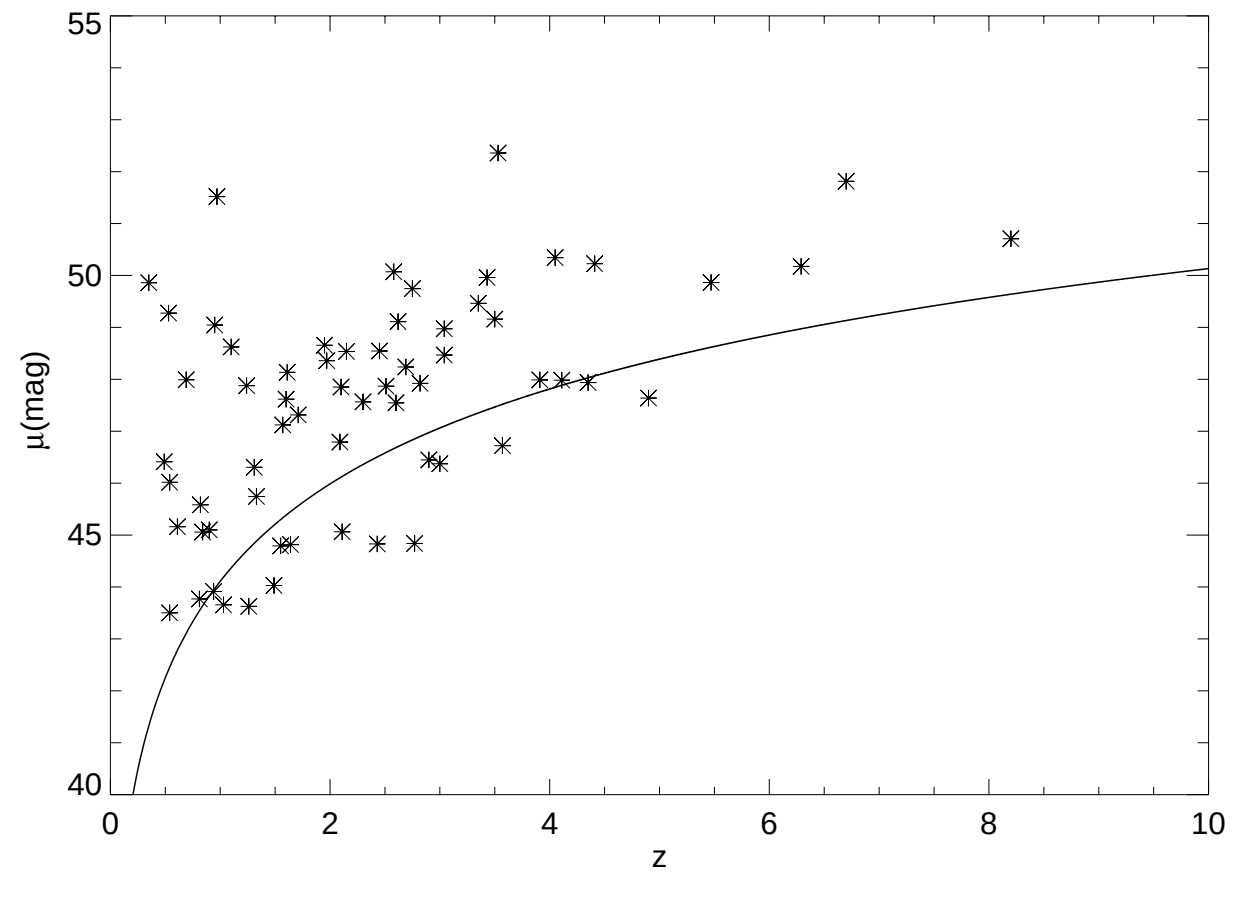

Fig. 5: The GRB Hubble diagram

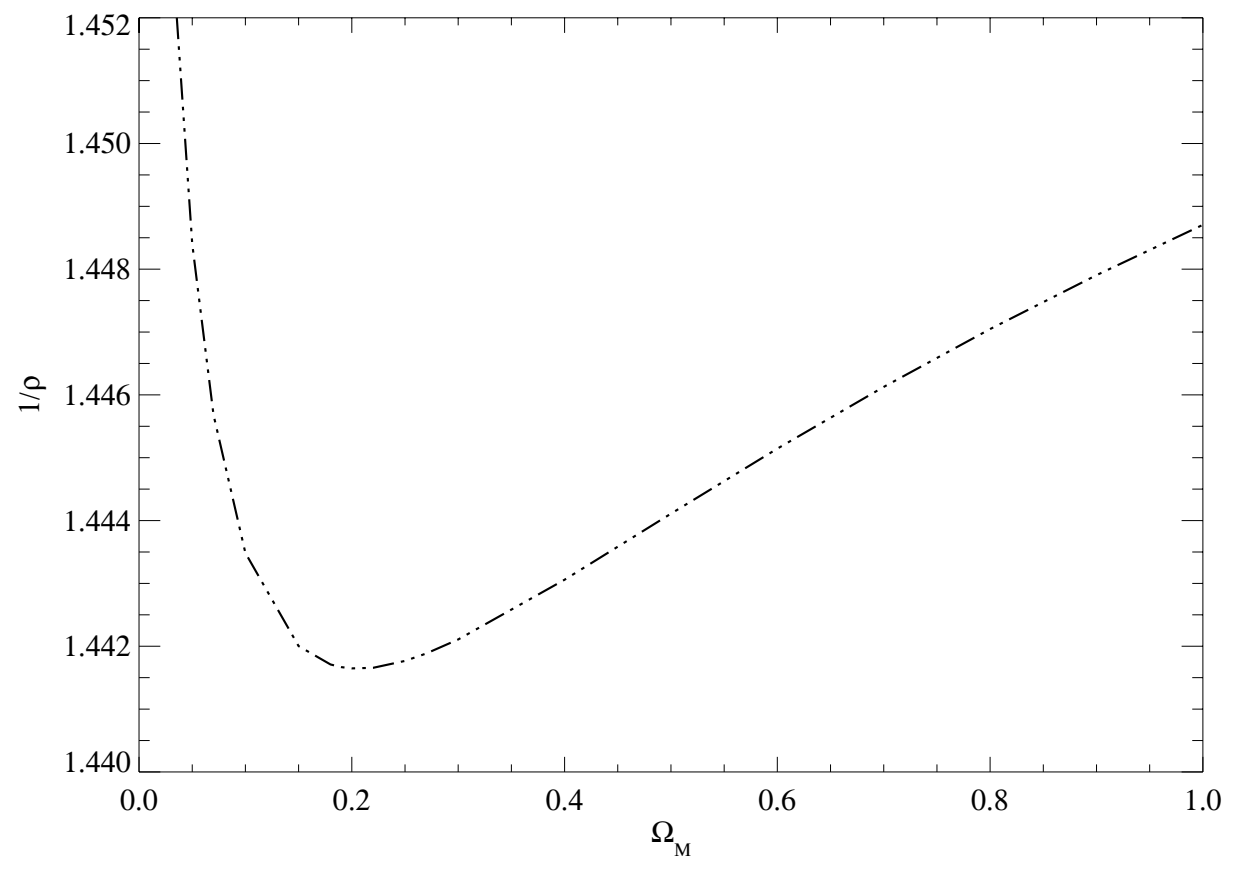

Fig. 6: A cosmological probe

$$\mu = 5\log_{10} D_L - 5 \tag{5}$$

It is seen that the points are mostly above the theoretical curve, so it is clear time-dilatation causes the same effect as in the case of supernovae. The final result is the most important one: an estimation of $\Omega_M$, the density parameter of matter. When the k-corrections were calculated, we had to give some input parameters, for example the mentioned quantity, because $D_L$ depends on the cosmology (see equation 6). It is clear that if we change these parameters, Amati's relation will also change.

$$D_L(z, \Omega_m, \Omega_\Lambda, H_0) = \frac{c(1+z)}{H_0} \cdot \frac{1}{\int_0^z \left[(1+z'^2)(1+\Omega_m z') - z'(2+z')\Omega_\Lambda\right]^{\frac{1}{2}} \, \mathrm{d}z'} \tag{6}$$

We can ask with what conditions we can get the highest correlation on the Amati plane. To find the answer, the calculations have been redone with so many values of $\Omega_M$, considering that $\Omega_{total} = 1$, i.e. a flat Universe, but there is a chance to test other models, too. We wanted to represent the correlation simply with the correlation coefficient $\rho$ (see Figure 6). This has a maximum when the correlation is highest, but all the results of some earlier papers [7] use methods where there is a minimum in the same case, and we therefore used the $1/\rho$ form in Figure 6. We can see that the minimum is at 0.2, which would be the optimal $\Omega_M$ for obtaining the highest correlation.

## 5 Conclusions

Our method, which is based on the k-correction, is a useful tool for cosmology. However, our estimation of $\Omega_M$ is less than the obtainable value from the WMAP data [8], which may be the most precise data that is available. Our results are approximately the same as the results of earlier GRB studies, although we have used different data.

### Acknowledgement

The Project is supported by the European Union, co-financed by the European Social Fund (grant agreement no. TAMOP 4.2.1./B-09/1/KMR-2010-0003), and in part through OTKA K077795, OTKA/NKTH A08-77719, and A08-77815 (Z.B.) grant.

András Kovács
Dept. of Physics of Complex Systems
Eötvös University
H-1117 Budapest, Pázmány P. s. 1/A, Hungary

Zsolt Bagoly
Dept. of Physics of Complex Systems
Eötvös University
H-1117 Budapest, Pázmány P. s. 1/A, Hungary

Lajos G. Balázs
Konkoly Observatory
H-1525 Budapest, POB 67, Hungary

István Horváth
Dept. of Physics
Bolyai Military University
Budapest, POB 15, H-1581, Hungary

Péter Veres
Dept. of Physics of Complex Systems
Eötvös University
H-1117 Budapest, Pázmány P. s. 1/A, Hungary
Dept. of Physics
Bolyai Military University
Budapest, POB 15, H-1581, Hungary